# Chemical Pressure effect at the boundary of Mott insulator and itinerant electron limit of Spinel Vanadates

P. Shahi[1], A. Kumar[1], Rahul Singh[1], Ripandeep Singh[2], P.U. Sastry,[2] A. Das[2], Amish G. Joshi[3], A. K. Ghosh[4], A. Banerjee[5] and Sandip Chatterjee[1,*]

[1]Department of Physics, Indian Institute of Technology (Banaras Hindu University), Varanasi-221 005, India
[2]Solid State Physics Division, Bhabha Atomic Research Center, Mumbai- 400085, India
[3]CSIR-National Physical Laboratory, Dr. K.S. Krishnan Road, New Delhi, India
[4]Department of Physics, Banaras Hindu University, Varanasi-221 005, India
[5]UGC-DAE Consortium for Scientific Research, Indore, Madhya Pradesh- 452017, India

ABSTRACT

The chemical pressure effect on the structural, transport, magnetic and electronic properties (by measuring X-ray photoemission spectroscopy) of $ZnV_2O_4$ has been investigated by doping Mn and Co on the Zinc site of $ZnV_2O_4$. With Mn doping the V-V distance increases and with Co doping it decreases. The resistivity and thermoelectric power data indicate that as the V-V distance decreases the system moves towards Quantum Phase Transition. The transport data also indicate that the conduction is due to the small polaron hopping. The chemical pressure shows the non-monotonous behaviour of charge gap and activation energy. The XPS study also supports the observation that with decrease of the V-V separation the system moves towards Quantum Phase Transition. On the other hand when Ti is doped on the V-site of $ZnV_2O_4$ the metal-metal distance decreases and at the same time the $T_N$ also increases.

INTRODUCTION

Mott insulators are important examples of strongly correlated materials. The strong-coupling limit, $U \gg t$ ($U$ is the inter-atomic Coulomb energy and $t$ is the spin dependent expectation value for the charge transfer between sites), corresponds to materials in which valence electrons are strongly localized in their atomic orbitals (Mott-Hubbard insulator). The opposite weak-coupling limit, $U \ll t$, corresponds to correlated metals whose electrons are completely delocalized (Paramagnetic metal). This implies that a Mott transition is induced at a critical value $Uc/t$ [1]. As a matter of fact, paramagnetic metals and Mott-Hubbard insulators represent two fundamentally different phases that can be interchanged by increasing or decreasing electronic correlations through a first-order quantum phase transition (QPT) [2]. It is highly challenging to characterize the electronic properties of materials when approaching the QPT either from Mott insulator side or from the paramagnetic metal side. Among very few materials, the $AV_2O_4$ spinels are a family of Mott insulators that fulfil the criterion of varying the $t/U$ ratio because of the metal-metal separation can be changed by applying the chemical pressure i.e., by changing the size of the $A^{2+}$ cation [3]. The absence of $e_g$ electrons makes direct V-V hybridization between $t_{2g}$ orbitals the only relevant contribution to the hopping amplitude. Moreover, $t$ is also a function of the interionic distance, $R$. This volume dependence between $d$ orbitals is $J/V^{-10/3}$ [4], which is the basis of the phenomenological Bloch's equation [5] for magnetic insulators: $\alpha = \partial \ln T_N / \partial \ln V = -3.3$ provided $U$ remains constant. When approached towards itinerant-electron behaviour the $\alpha$ value is very sensitive so that it can indicate the applicability of crystal-field theory [6,7].

$ZnV_2O_4$ has one of the smallest charge gaps among the vanadium spinels: ($\Delta \approx 0.55$ eV) [8,9]. Therefore, with proper doping on A-site this $ZnV_2O_4$ may be transferred to the Mott transition. In a recent report [10] a clear evidence of pressure dependence of a nonmonotonic behaviour of charge gap was shown. The nonmonotonic dependence of charge gap on $t/U$ is expected for materials that are deep inside the Mott regime (small $t/U$ ratio) [11,12] and the enhancement of the magnetic up-up-down-down ordering along the [±1,0,1] and [0,±1,1] chains has already been reported in these vanadates[13].

$ZnV_2O_4$ crystallizes in a cubic spinel structure, where the V atoms form a pyrochlore lattice of corner-sharing tetrahedra. As a matter of fact, the antiferromagnetic (AF) interactions between the V atoms are highly frustrated. At $T_S = 51$ K, $ZnV_2O_4$ undergoes a structural phase transition where the symmetry is lowered from cubic to tetragonal (c/a < 1) with a compression of the $VO_6$ octahedron along the $c$ axis [14] and the system possibly orbital orders. This also lifts the geometrical frustration of the cubic phase making way for the second transition at $T_N = 40$ K which is of a magnetic nature and the system orders antiferromagnetically [14-16]. The lattice formed by the V atoms can be described as built up by V-V chains running along the [110], [011], and [101] directions. The magnetic structure, found by neutron diffraction [13,14], is AF along the [110] direction (within the *ab* plane), but along the [101] and [011] (off plane) directions the V moments order. Moreover, the effective hopping and the *d-d* transfer integral of $ZnV_2O_4$ estimated from the X-ray photoemission spectra are close to the metallic $LiV_2O_4$ [17].

Moreover, it has been shown that in $ZnV_2O_4$ away from the strong-coupling regime *bond and magnetic* ordering are independent to each other [10]. If the bond ordering is strengthened away from the strong-coupling regime then it may correspond to ferroelectricity as is observed in $CdV_2O_4$ [18]. It has also been suggested by Kuntscher et al. [10] that ferroelectricity can be increased significantly by decreasing V-V distance as the structural distortion observed in $ZnV_2O_4$ is identical to $CdV_2O_4$. Furthermore, the main limiting factors for $ZnV_2O_4$ to show spontaneous polarization are the losses or leak currents. The increase in the value of charge gap reduces the losses. In $ZnV_2O_4$ when external pressure is applied, which lead to the decrease in V-V distance, the charge gap increases above a critical pressure [10]. As the V-V separation decreases the system moves towards the itinerant electron limit [Blanco-Canasa]. Therefore, closing the Mott gap of this geometrically frustrated Mott insulator is a potential route for enhancing ferroelectricity which could be achieved in $ZnV_2O_4$ by reducing the V-V distance. Chemical substitution is the best way in doing this.

In the present paper we have varied the chemical pressure in $ZnV_2O_4$ by doping Mn and Co on the Zn site and studied the effect of it on magnetic, optical, transport properties and on electron structure of doped and undoped $ZnV_2O_4$. We have also doped Ti on the V site and varied the metal-metal distance and observed its effect. Our results support the universality of the first order character of the transition from the localized to the itinerant-electron and the stabilization of an intermediate phase in which there is a disproportionation into molecular orbitals within clusters and weak bonding between clusters [19,20].

EXPERIMENT

Polycrystalline $(Zn_{1-x}Mn_x)V_2O_4$, $Zn_{0.9}Co_{0.1}V_2O_4$ and $Zn(V_{1-x}Ti_x)_2O_4$ (where x = 0, 0.05 and 0.10) were prepared by solid state reaction route. Appropriate ratio of ZnO, CoO, $V_2O_3$, $Ti_2O_3$ and MnO were mixed and pressed into pellets. The pellets were sealed in quartz tube under high vacuum and heat treated into furnace at 800°C for 60 hrs. The X-ray diffraction measurement was performed at 10K and at room temperature. The magnetic measurements were done with Vibrating Sample magnetometer (VSM). Fourier Transform Infrared Spectroscopy measurements have been done using Spectrum 65 FTIR spectrometer (Perkin Elmer Instruments, USA) in the range of 500 to 4000 cm$^{-1}$. Resistivity measurements has been performed by four probe method and thermoelectric power measurement (Seebeck coefficient) has been done using home-made thermoelectric setup. X-ray Photoelectron Spectroscopy (XPS) experiments were performed using Omicron Nanotechnology UHV system equipped with a twin anode Mg/Al X-ray source (DAR400), a monochromator and a hemispherical electron energy analyzer (EA125 HR). All the XPS measurements were performed inside the analysis chamber under base vacuum of 5.0x10$^{-10}$ Torr using monochromatized AlKα. The total energy resolution, estimated from the width of the Fermi energy, was about 300 meV for monochromatic AlKα line with photon energy 1486.60 eV.

RESULTS & DISCUSSION

The XRD measurement of all the samples indicates the single phase nature of the samples. The samples crystallize in cubic phases with Fd-3m space group. At low temperature (structural transition temperature) the structure changes to tetragonal phase with I41/a space group. We have refined the diffraction data using the Fullprof program (shown in Fig.1) and the fitted parameters are shown in Table-I. In the inset of Fig. 1 the X-ray diffraction pattern of Mn and Ti doped samples have been shown in a very narrow 2θ (61.5°-63°) range at 300K (above the structural transition) and at 10K (below the structural transition, if any). A very weak splitting is observed for 10% Mn and Co doped samples in the 10K diffraction pattern indicating the decrease of structural distortion. Moreover, for Ti doped samples also structural transition is reduced. The estimated c/a ratio for undoped sample is 0.9892, whereas, for 10% Mn, Co and Ti doped samples it is ~0.9894, ~0.9897 and ~0.9893 respectively. Magnetization curves of $Zn_{1-x}A_xV_2O_4$ [A=Mn and Co] measured for different doping concentration are shown in Fig.2. For $ZnV_2O_4$ two clear transitions are observed where the higher temperature is the structural transition and the lower temperature is the magnetic transition [14-16]. It is observed that with increase of doping concentration the peak appears due to structural transition is reduced which is consistent with the XRD measurement.

Figure 3 shows the evolution of $T_N$ with the V-V separation for the $Zn_{1-x}A_xV_2O_4$ [A=Mn and Co] spinels. It is to be mentioned that as the overlap integral in t of J ∞ $t^2$/U [J is the super exchange spin-spin interaction] increases with decreasing $R_{V-V}$, the energy U must decrease or remain constant. As a matter of fact, within the localized-electron superexchange, $T_N$ should increase with decreasing V-V separation. In the present investigation it has been observed when Mn is doped the V-V separation increases and subsequently $T_N$ decreases. On the other hand, when Co is doped it is observed that V-V separation decreases very slightly, but interestingly $T_N$ decreases. This suggests that the energy U in the localized electron superexchange theory is not constant when Co is doped.

As mentioned above, the variation of $T_N$ with V-V distance can be interpreted as the consequence of a variation of U/t, which collapses close to the itinerant limit. According to this, even though $ZnV_2O_4$ and Co-doped $ZnV_2O_4$ are still semiconducting, a partial electronic delocalization along the V-V bonds may occur as has been anticipated by Pardo et al. [8] for their $ZnV_2O_4$ and $MgV_2O_4$ samples which are at the boundary between itinerant and localization limits. In Fig.4 we have shown the experimental resistivity curves for Mn and Co doped and undoped $ZnV_2O_4$ in which the V-V distances are different. The results show the activation energy decreases with the decrease of V-V distance as the metal-insulator transition is approached from the localized-electron side which confirms the reduction of U/t with decrease of V-V distance, which might be due to a partial delocalization of electrons [3] and in this limit neither the localization-electron model and nor the itinerant-electron model is applicable as has been mentioned by Blanco-canosa et al [3]. It is observed from the resistivity curve at low temperature with the decrease of V-V distances the $V^{3+}$-3d electrons approaches to the Quantum Phase Transition (QPT) from the localized-electron side. In Fig. 5 the temperature dependence of the thermoelectric power, S(T) is shown for all the spinels mentioned above. The S(T) value is large and positive which indicates that there is only a low density of mobile holes in these vanadates. Thermally activated behaviour of small polarons for all these vanadates is observed. Therefore, the possibility of the existence of the large polaron which forms due to non-stoichiometry can be excluded. Also, the S value decreases from Mn-doped to undoped to the Co-doped $ZnV_2O_4$ sample.

Moreover, it is observed that $ZnV_2O_4$ and Co doped $ZnV_2O_4$ exhibit low activation energy compared to the Mn doped $ZnV_2O_4$ samples. In consistent with the earlier study [3] a structural transition from the low temperature tetragonal to a high temperature cubic phase is observed. These transitions are due to a co-operative ordering of strong V-V bonding as is observed from XRD refinement. It is found for Mn doped samples the V-V bonding becomes weaker and it moves towards the localized electron side. As is already mentioned that the estimated phenomenological parameter is unfeasible in the Mn-doped samples, therefore, U>>t might be the case in this Mn doped samples which are in localized electron side.

Furthermore, it is observed in Fig. 2 the susceptibility drops sharply for $ZnV_2O_4$ and Co-doped $ZnV_2O_4$ on cooling through the structural transition temperature. This is due to the spin-pairing of V-V bonds [3]. For Mn doped samples this kind of sharp drop is not observed. Furthermore, from the $d\chi/dT$ vs T plot [Fig. 6] it is observed that with increase of Mn content the high temperature peak which is associated with the structural transition is diminished. This behaviour i.e. the suppression of the orbital ordering when it is going towards the Mott-regime (t<<U) is consistent with those already reported [11,12].

We have also measured the transmittance of all the samples and plotted the absorbance as a function of frequency in Fig. 7. A peak is observed below 1000 $cm^{-1}$ which is due to one of the four phonon modes as is observed for cubic spinels. The absorbance spectra show a strong onset above 1000 $cm^{-1}$. The variation of charge gap ($\Delta$), estimated from the intersection of the linear extrapolation of the absorption edge and the frequency axis, with the inverse of V-V distance is shown in Fig. 8. The $\Delta$ initially decreases with increase of V-V distance and with further increase of V-V distance it increases sharply. This non-monotonous behaviour is consistent with that of the external pressure effect [10]. This result we have confirmed by plotting the activation energy (estimated from the resistivity measurement) as a function of inverse V-V distance [Fig. 8]. We have seen that the variation of activation energy with chemical pressure is consistent with that of charge gap variation with chemical pressure. It has already been proposed [8] that in $ZnV_2O_4$ the V-V dimmers form along the [011] and [101] directions together with an up-up-down-down magnetic order along the same directions. Moreover, in chemical pressure dependence of sample volume [Fig.9] an anomaly is observed for the V-V distance 2.9736 Å. This is consistent with the inverse V-V distance variation of charge gap data. It might be the fact that stabilization of dimerized phase occurs below the V-V distance 2.9736 Å. Similar kind of behaviour is observed when 10-12 GPa pressure is applied on $ZnV_2O_4$ system. Therefore, it might be predicted that 10-12 GPa pressure reduces the V-V distance to 2.9736Å. Therefore, we see from the above discussion that $T_N$, charge gap and activation energy of different spinel vanadates depends on the V-V separation and as it decreases a breakdown of localized electron model for the 3d electrons is observed. This observation is consistent with those observed previously [3,10]. The shortest V-V distance we have obtained in Co doped $ZnV_2O_4$ and that distance is 2.9738 Å which is larger than the critical distance for electron itineracy (2.84Å) [20]. This is similar to the case of $MgTi_2O_4$ where a tetramerization of the Ti chains is observed [21].

We have also studied the electronic structure of Mn and Co doped $ZnV_2O_4$ using x-ray photoemission spectroscopy (XPS). The purpose of this study is to see what happens in the electronic structures when the system is moving from itinerant electron side to localized electron side and vice versa and to provide clues to understand the origin of charge and orbital orderings in the spinel-type V oxides with tetragonal distortions. Zn 2$p$, V 2$p$, Mn 2$p$ and Co 2$p$ XPS core-level spectra of ZnV2O4, and Mn and Co doped ZnV2O4 are shown in Fig. 10. All the core level spectra show 2$p^{3/2}$ and V 2$p^{1/2}$ states. Both Mn 2$p$ and Co 2$p$ spectra show satellites peaks. The two clear distinct states of V 2$p^{3/2}$ and V 2$p^{1/2}$ observed at ~7.6 eV apart, due to spin-orbit splitting. The V 2$p^{3/2}$ spectrum of ZnV$_2$O$_4$ indicates a V$^{3+}$ single valence level and no indication of the existence of V$^{4+}$ (or V$^{5+}$) component as has been observed in some other reports [22-24].

Three features (A, B and C) are observed in the valence-band spectra for ZnV$_2$O$_4$, and Mn and Co doped ZnV$_2$O$_4$ (Fig.11). Feature (A) close to Fermi level ($E_F$) can be assigned to the antibonding band of the V 3$d$ $t_{2g}$ states hybridized with the O 2$p$ states, while features B and C are due to the bonding band of the O 2$p$ states hybridized with the V 3$d$ $t_{2g}$ and $e_g$ states, respectively [25, 26]. The position of A shifts to the $E_F$ in going from Mn doped ZnV$_2$O$_4$ to ZnV$_2$O$_4$ to Co doped ZnV$_2$O$_4$ and, consequently, the magnitude of the band gap in $t_{2g}$ bands decreases in this order. The magnitude of the band gap for ZnV$_2$O$_4$ ~0.2 eV which is in good agreement with that estimated from the resistivity data. In the case of Co doped ZnV$_2$O$_4$, the V 3$d$ band reaches very close to $E_F$, consistent with the resistivity behavior. The binding energy of feature B for Co doped ZnV$_2$O$_4$ (4.79 eV), corresponding to the $p$-$d$ charge transfer energy, is small compared to those for ZnV$_2$O$_4$ and Mn doped ZnV$_2$O$_4$ (~5.32 eV).

Takubo *et al.* [17] calculated the tight-binding (TB) parameters for ZnV$_2$O$_4$ and CdV$_2$O$_4$ using Harrison's rule [27], where the parameters were defined as $(pd\sigma)\alpha 1/(R_{V-O})^{3.5}$, $(dd\sigma)\alpha 1/(R_{V-V})^5$, $(pd\sigma)/(pd\pi)=(dd\sigma)/(dd\pi)=-2.16$ and $R_{V-O}$ and $R_{V-V}$ are the bond lengths between the neighboring V and O sites and between the neighboring two V sites, respectively. The set for the structural parameters and TB parameters are listed in Table II. The $pd\pi$ value for Co doped ZnV$_2$O$_4$ (~0.95 eV) is slightly larger than that for ZnV$_2$O$_4$ (0.94eV) and Mn doped ZnV$_2$O$_4$ (0.93eV). Also it is observed that the $dd\sigma$ value increases slightly in going from Co doped ZnV$_2$O$_4$ (−0.52 eV) to ZnV$_2$O$_4$ (−0.47 eV) to Mn doped ZnV$_2$O$_4$ (−0.40 eV). As a result, the effective hopping due to $d$-$d$ transfer $T\sigma=3/4(dd\sigma)$ is ranging from −0.350 eV to −0.356 eV, while the effective hopping due to $p$-$d$ transfer $T\pi=-(pd\pi)^2/\Delta$ is ranging from −0.185 eV to −0.21 eV. Here, the values of $\Delta$ are estimated from the XPS spectra. This indicates that the exchange interaction due to $d$-$d$ transfer is dominant in the systems. The magnitude of $T_\sigma$ gradually decreases in going from Co doped ZnV$_2$O$_4$ to ZnV$_2$O$_4$ to Mn doped ZnV$_2$O$_4$. The above results show that variations of different parameters with doping are not so significant. But the tendency clearly indicates that Co-doped $ZnV_2O_4$ are closer to the itinerant electron limit. Takubo et al.[17] has also shown that for metallic $LiV_2O_4$ the *pdπ value is larger but ddσ* value is smaller compared to the semiconducting $ZnV_2O_4$ and $CdV_2O_4$. This also supports our observation.

Takubo *et al.* [17] have explained the XPS for $ZnV_2O_4$ and $CdV_2O_4$ by orbital-driven Pierls (ODP) and complex linear combination orbitals (COO) model [28,29]. The ODP scenario is based on the assumptions that the $t_{2g}$ band has one-dimensional character due to the dominant *d-d* direct hopping and that the $t_{2g}$ band has itinerant character due to the closeness to the metal-insulator transition. The present photoemission study shows the large $dd\sigma$ values indicating the the dominant *d-d* direct hopping and this increases from Mn doped $ZnV_2O_4$ to Co doped $ZnV_2O_4$ through $ZnV_2O_4$. Moreover from the above discussion on valence band spectra it is observed that V 3d band reaches closer to the $E_F$ from Mn doped to Co doped $ZnV_2O_4$ through $ZnV_2O_4$. Therefore, when it is going towards the smaller $R_{V-V}$ the Pierls Physics is becoming important over Mott Physics.

In the present investigation we have also doped Ti on the Vanadium site of $ZnV_2O_4$ to vary the metal-metal distance. It is observed from the refinement of the XRD data that with increase of Ti content the metal-metal distance increases. On the other hand, with increase of Ti content the $T_N$, charge gap and the activation energy also increase [Figs.12, 13 and 14]. As is already mentioned with localization electron model with increase of metal-metal distance the $T_N$ should decrease. But in the present case as Ti is doped opposite behaviour is observed. It may happen that Ti-doping decreases the intra-atomic Coulomb energy U which in effect increases the $T_N$. Nevertheless, it deserves further study to explain these interesting phenomena. Moreover, $MgTi_2O_4$ undergoes a M-I transition on cooling below $T_{MI}$ ~ 260 K, accompanied by a strong decrease of the magnetic susceptibility and a transition to a tetragonal structure [30]. The structure in this composition was found to contain dimmers with short Ti-Ti distances (2.85 Å), the locations of the spin singlets. The electronic structure is consistent with the opening of a 1 eV gap and the absence of magnetic moments and enables one to interpret the crystal structure in terms of orbital ordering [20]. But in the present case when Ti is doped the metal-metal distances increase. That might be reason of interesting behaviour observed in this Ti doped samples.

CONCLUSION:

The magnetization data show that when 10% Mn and Co is doped in $ZnV_2O_4$ the intensity of peak due to the structural transition goes down. When Mn is doped the resistivity and thermoelectric power increase along with the increase of V-V distance. when Co is doped the V-V separation decreases very slightly, but $T_N$ decreases which suggests that the interatomic Coulomb energy U in the localized electron superexchange theory is not constant when Co is doped. The activation energy decreases with the decrease of V-V distance as the metal-insulator transition is approached from the localized-electron side which confirms the reduction of U/t with decrease of V-V distance, which might be due to a partial delocalization of electrons and in this limit neither the localization-electron model and nor the itinerant-electron model is applicable. It is observed from the resistivity curve that at low temperature as the V-V distance decreases the $V^{3+}$-3d electrons approaches to the Quantum Phase Transition (QPT) from the localized-electron side. Thermally activated behaviour of small polarons for all these vanadates is observed from temperature variation of thermoelectric data. From the variation of $d\chi/dT$ with T plot it is observed that with increase of Mn, that is, with the increase of chemical pressure structural transition is suppressed. Along with $T_N$, also the charge gap and activation energy of the vanadates depend on the V-V separation and as it decreases a breakdown of localized electron model for the 3d electrons is observed. The XPS study also indicates that with decreasing V-V separation the system moves towards the itinerant electron limit. Ti-doped $ZnV_2O_4$ samples show some un-usual behaviour. With Ti doping metal-metal distances increase but at the same time $T_N$ increases. It deserves further study to explain the actual mechanism in this Ti doped sample.

**Acknowledgement**

SC is grateful to DST (Grant No.: SR/S2/CMP-26/2008), CSIR (Grant No.: 03(1142)/09/EMR-II) and BRNS, DAE ((Grant No.: 2013/37P/43/BRNS) for providing financial support. The authors are also grateful to UGC-DAE Consortium for Scientific Research, Indore, India for providing facility for magnetization measurement.

**Table 1** Structural parameters (lattice parameters, bond lengths) of $Zn_{1-x}A_xV_2O_4$ (with x=0, 0.05, 0.1 and A=Mn, Co, Ti)) samples obtained from Reitveld refinement. The structural data have been refined with Space group I41/a at10K and Fd-3 m at 300K.

| | | | | | |
|---|---|---|---|---|---|
| $ZnV_2O_4$ | 10 K | a(Å) | 5.9481(1) | d(Zn-O)(Å) | 4 x 1.96874(4) |
| | | b(Å) | 5.9481(1) | d(V-O) (Å) | 2×2.00193(7), 4×2.02377(3) |
| | | c(Å) | 8.3713(4) | d(V-V)(Å) | 2 x 2.96689(6), 4 x 2.97406(7) |
| | 300 K | a(Å) | 8.4130(1) | d(Zn-O)(Å) | 4 x 1.974774 (3) |
| | | | | d(V-O) (Å) | 6 x 2.01864(3) |
| | | | | d(V-V)(Å) | 6 x 2.97446(3) |
| $Zn_{0.95}Mn_{0.05}V_2O_4$ | 10 K | a(Å) | 5.9503(3) | d(Zn-O)(Å) | 4 x 1.970(6) |
| | | b(Å) | 5.9503(3) | d(V-O) (Å) | 2 x 2.003(7), 4 x 2.024(4) |
| | | c(Å) | 8.3769(2) | d(V-V)(Å) | 2 x 2.97517(8), 4 x 2.96843(6) |
| | 300 K | a(Å) | 8.4155 (2) | d(Zn-O)(Å) | 4 x 1.975359(3) |
| | | | | d(V-O) (Å) | 6 x 2.01924(2) |
| | | | | d(V-V)(Å) | 6 x 2.97534(2) |
| $Zn_{0.9}Mn_{0.1}V_2O_4$ | 10 K | a(Å) | 5.9540(4) | d(Zn-O)(Å) | 4 x 1.97083(8) |
| | | b(Å) | 5.9540(4) | d(V-O) (Å) | 2 x 2.00432(12), 4 x 2.02578(7) |
| | | c(Å) | 8.3813(1) | d(V-V)(Å) | 2 x 2.97700(15), 4 x 2.97013(10) |
| | 300 K | a(Å) | 8.4174 (1) | d(Zn-O)(Å) | 4 x 1.97579(2) |
| | | | | d(V-O) (Å) | 6 x 2.01968(4) |
| | | | | d(V-V)(Å) | 6 x 2.97599(5) |
| $Zn(V_{0.95}Ti_{0.05})_2O_4$ | 10 K | a(Å) | 5.9487(2) | d(Zn-O)(Å) | 4 x 1.96893(5) |
| | | b(Å) | 5.9487(2) | d(V-O) (Å) | 2 x 2.00210(10), 4 x 2.02398(5) |
| | | c(Å) | 8.3719(7) | d(V-V)(Å) | 2 x 2.97436(10), 4 x 2.96716(8) |
| | 300 K | a(Å) | 8.4115 (6) | d(Zn-O)(Å) | 4 x 1.97442(3) |
| | | | | d(V-O) (Å) | 6 x 2.01828(5) |
| | | | | d(V-V)(Å) | 6 x 2.97393(5) |
| $Zn(V_{0.9}Ti_{0.1})_2O_4$ | 10 K | a(Å) | 5.9487(2) | d(Zn-O)(Å) | 4 x 1.96898(8) |
| | | b(Å) | 5.9487(2) | d(V-O) (Å) | 2 x 2.00227(12), 4 x 2.02398(7) |
| | | c(Å) | 8.3726(8) | d(V-V)(Å) | 2 x 2.96728(10), 4 x 2.97435(15) |
| | 300 K | a(Å) | 8.4102 (7) | d(Zn-O)(Å) | 4 x 1.97411(3) |
| | | | | d(V-O) (Å) | 6 x 2.01796(5) |
| | | | | d(V-V)(Å) | 6 x 2.97345(5) |
| $Zn_{0.9}Co_{0.1}V_2O_4$ | 10 K | a(Å) | 5.9466(2) | d(Zn-O)(Å) | 4 x 1.96860(4) |
| | | b(Å) | 5.9466(2) | d(V-O) (Å) | 2 x 2.00248(7), 4 x 2.02327(4) |
| | | c(Å) | 8.3736(1) | d(V-V)(Å) | 2 x 2.97331(8), 4 x 2.96692(6) |
| | 300 K | a(Å) | 8.4111(9) | d(Zn-O)(Å) | 4 x 1.97431(2) |
| | | | | d(V-O) (Å) | 6 x 2.01817(4) |
| | | | | d(V-V)(Å) | 6 x 2.97377(4) |

**Table 2** Tight-binding parameters for $ZnV_2O_4$, $Zn_{0.9}Mn_{0.1}V_2O_4$ and $Zn_{0.9}Co_{0.1}V_2O_4$.

| Sample | Phase | Distance($A^0$) V-V | V-O | $dd\sigma$ | $pd\pi$ | $\Delta$(ev) | $T_\sigma$ | $T_\pi$ |
|---|---|---|---|---|---|---|---|---|
| $ZnV_2O_4$ | Cubic | 2.97446 | 2.01864 | -0.4639 | 0.9321 | 4.89 | -0.3479 | -0.1778 |
| | Tetragonal | 2.96689, 2.97406 | 2.02377, 2.00193 | -0.4742, -0.4685 | 0.9305, 0.9666 | | -0.3557, -0.3514 | -0.1771, -0.1911 |
| $Zn_{0.9}Mn_{0.1}V_2O_4$ | Cubic | 2.97599 | 2.01968 | -0.4627 | 0.9304 | 5.32 | -0.3470 | -0.1627 |
| | Tetragonal | 2.97013, 2.97700 | 2.02578, 2.00432 | -0.4716, -0.4662 | 0.9237, 0.9625 | | -0.3537, -0.3497 | -0.1604, -0.1741 |
| $Zn_{0.9}Co_{0.1}V_2O_4$ | Cubic | 2.97377 | 2.01817 | -0.4645 | 0.9329 | 4.79 | -0.3484 | -0.1817 |
| | Tetragonal | 2.96692, 2.97331 | 2.02327, 2.00248 | -0.4741, -0.4691 | 0.9313, 0.9656 | | -0.3556, -0.3518 | -0.1811, -0.1947 |

**Figure Captions:**

1. X-ray diffraction pattern with Reitveld refinement for Mn, Co and Ti doped $ZnV_2O_4$ samples at (a) 300K and (b) 10K. (c) X-ray diffraction pattern in very narrow $2\theta$ (61.5°-63°) range at 300K (above the structural transition) and at 10K.

2. Temperature variation of magnetization for $Zn_{1-x}A_xV_2O_4$ [where x=0.05 and 0.1 for A=Mn and x =0.01 for A = Co] spinels.at H=500 Oe.

3. Variation of $T_N$ with the inverse V-V distance, $1/R_{V-V}$ for the $Zn_{1-x}A_xV_2O_4$ [where x=0.05 and 0.1 for A=Mn and x =0.01 for A = Co] spinels.

4. Temperature dependent Resistivity curves for Mn and Co doped $ZnV_2O_4$.

5. Temperature dependent of thermoelectric power for $Zn_{1-x}A_xV_2O_4$ [where x=0.05 for A=Mn and x =0.01 for A = Co].

6. $d\chi/dT$ as a function of Temperature for Mn and Co doped $ZnV_2O_4$.

7. Room temperature Absorbance Spectra of $Zn_{1-x}A_xV_2O_4$ [where x=0.05 and 0.1 for A=Mn and x =0.01 for A = Co] and $Zn(V_{1-x}Ti_x)_2O_4$ [where x=0.05 and 0.1].

8. Charge gap and activation energy variation as a function of inverse V-V distance for $Zn_{1-x}A_xV_2O_4$ [where x=0.05 and 0.1 for A=Mn and x =0.01 for A = Co].

9. Variation of Volume as a function of $1/R_{V-V}$ for $Zn_{1-x}A_xV_2O_4$ [where x=0.05 and 0.1 for A=Mn and x =0.01 for A = Co].

10. V 2p XPS core-level spectra of $ZnV_2O_4$ (a), $Zn_{0.9}Mn_{0.1}V_2O_4$ (b) and $Zn_{0.9}Co_{0.1}V_2O_4$ (C); Zn 2p XPS core-level spectra of $ZnV_2O_4$ (d), $Zn_{0.9}Mn_{0.1}V_2O_4$ (e) and $Zn_{0.9}Co_{0.1}V_2O_4$ (f); Mn 2p XPS core-level spectra of $Zn_{0.9}Mn_{0.1}V_2O_4$ (g); Co 2p XPS core-level spectra of $Zn_{0.9}Co_{0.1}V_2O_4$ (h).

11. Valance-band XPS spectra of $ZnV_2O_4$, $Zn_{0.9}Mn_{0.1}V_2O_4$ and $Zn_{0.9}Co_{0.1}V_2O_4$.

12. Temperature variation of magnetization measured at H=500 Oe for $Zn(V_{1-x}Ti_x)_2O_4$ [where x=0.05 and 0.1].

13. Temperature dependent Resistivity and Seeback Coefficient for $Zn(V_{1-x}Ti_x)_2O_4$ [where x=0.05 and 0.1].

14. Variation of Charge gap and Activation energy as a function of $1/R_{V-V}$ for $Zn(V_{1-x}Ti_x)_2O_4$ [where x=0.05 and 0.1].

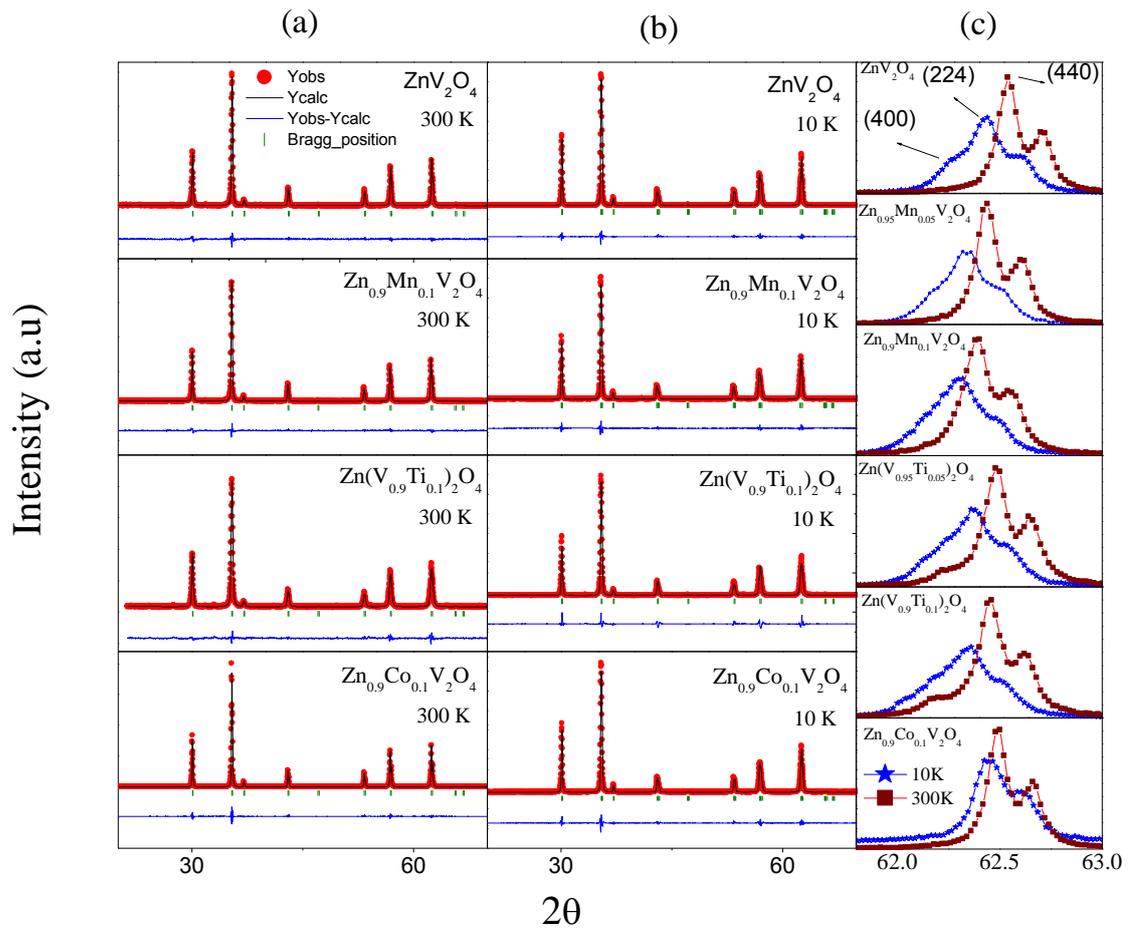

Figure1

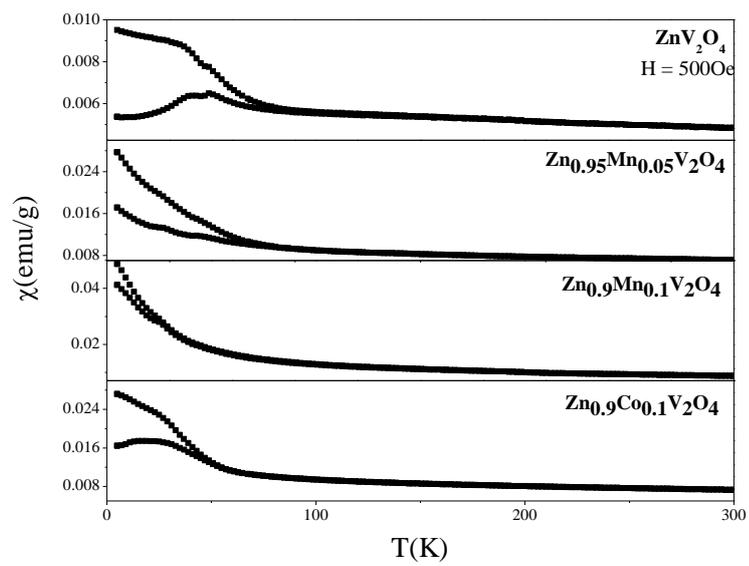

Figure 2

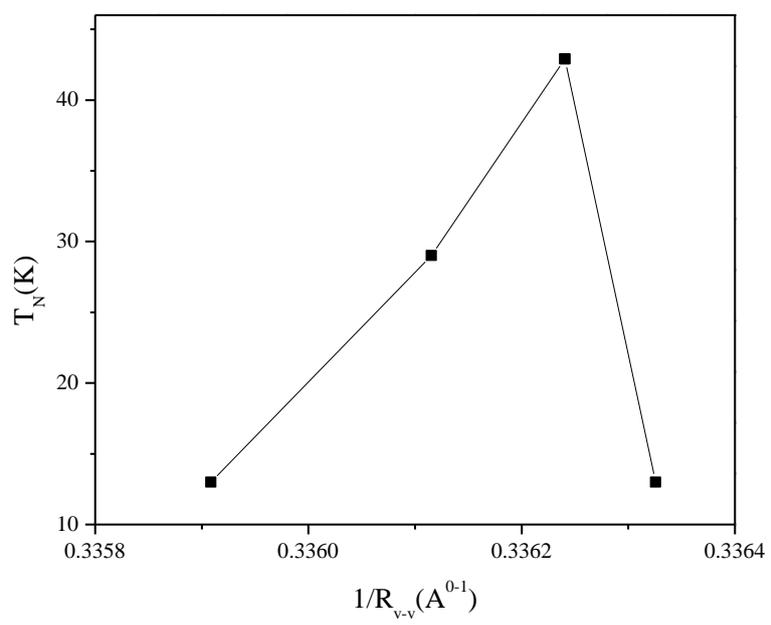

Figure 3

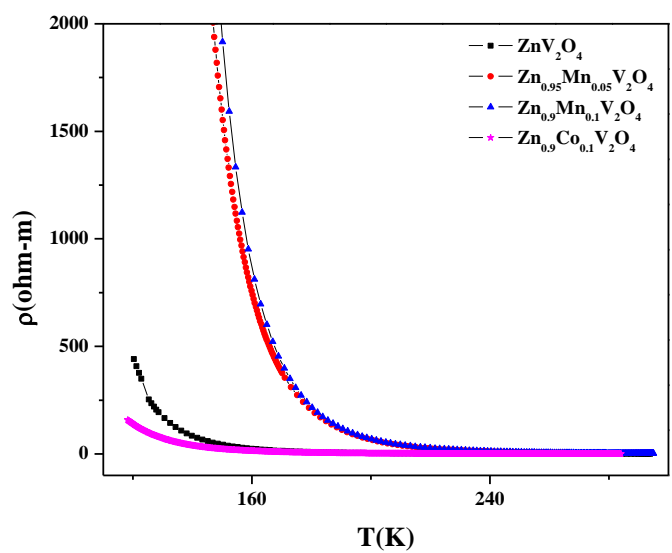

Figure 4

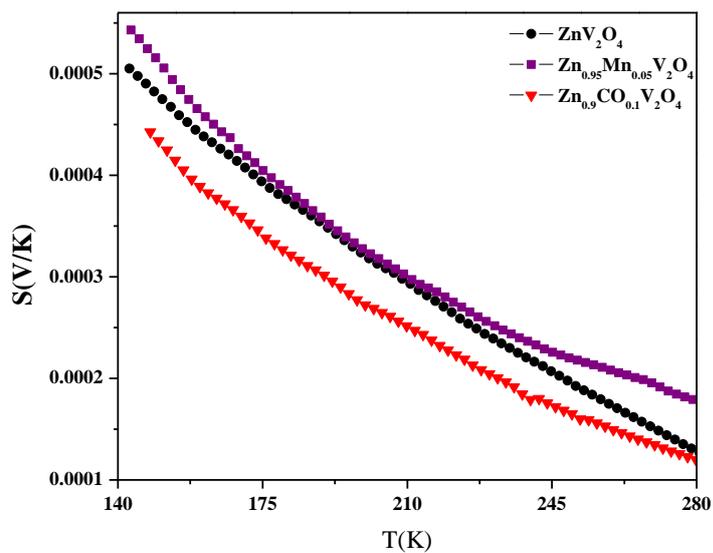

Figure 5

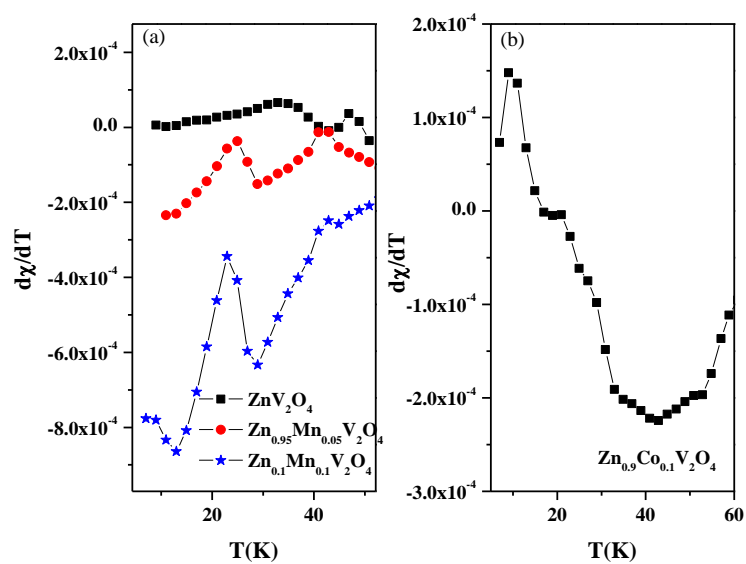

Figure 6

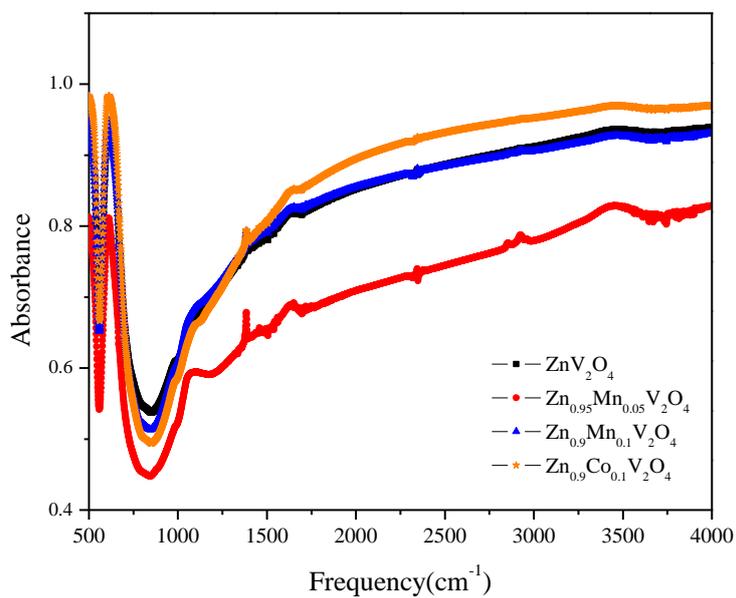

Figure 7

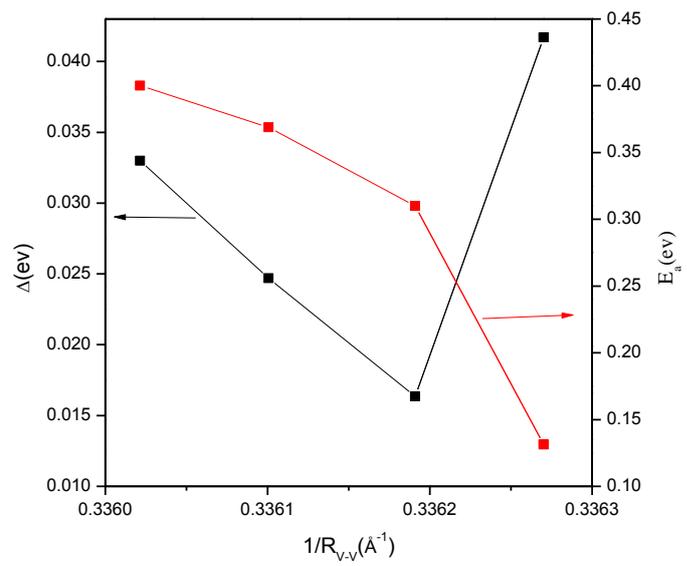

Figure 8

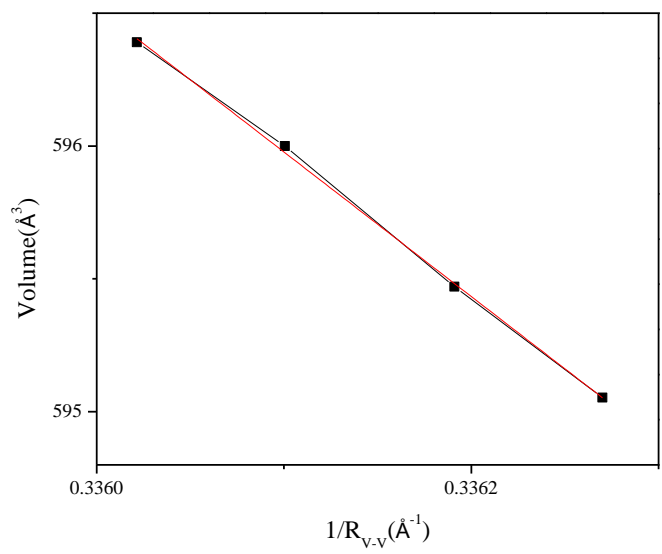

Figure 9

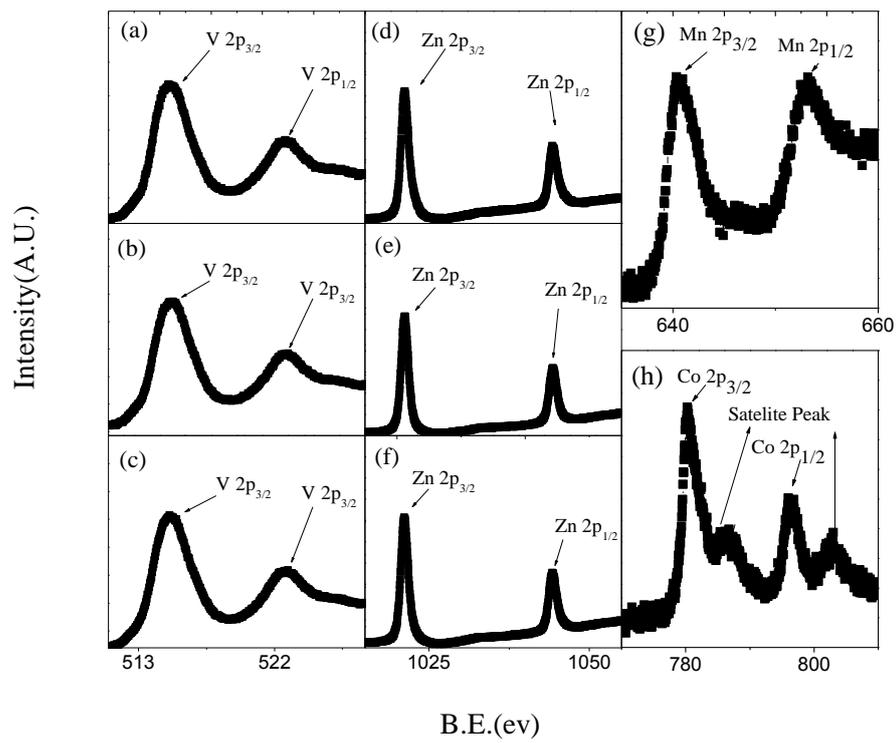

Figure 10

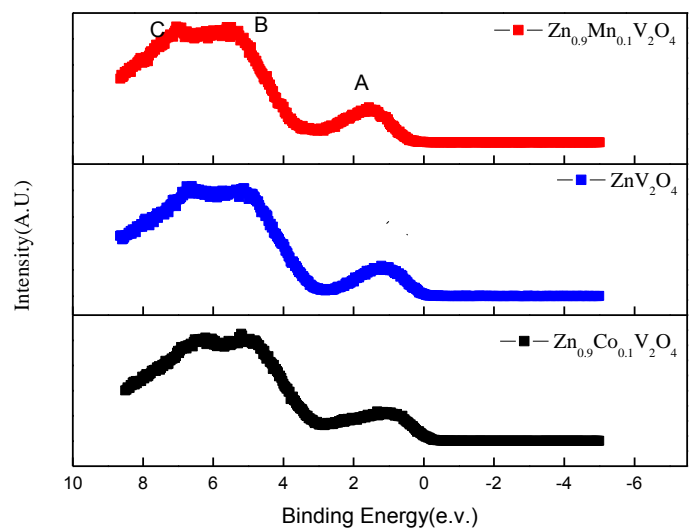

Figure 11

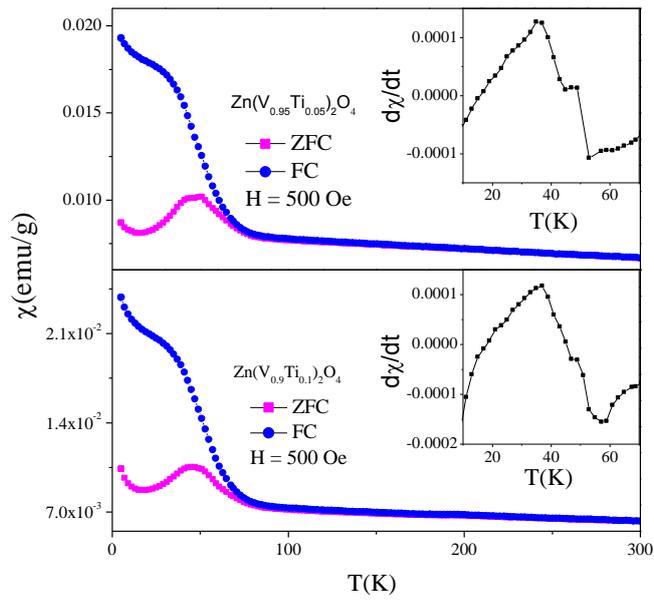

Figure12

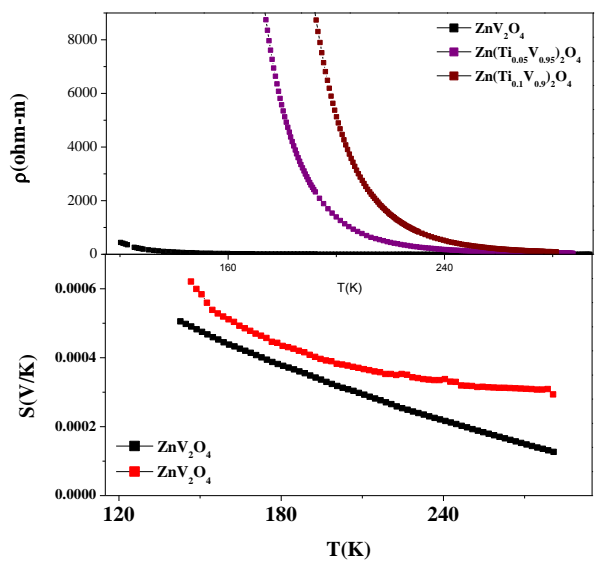

Figure 13

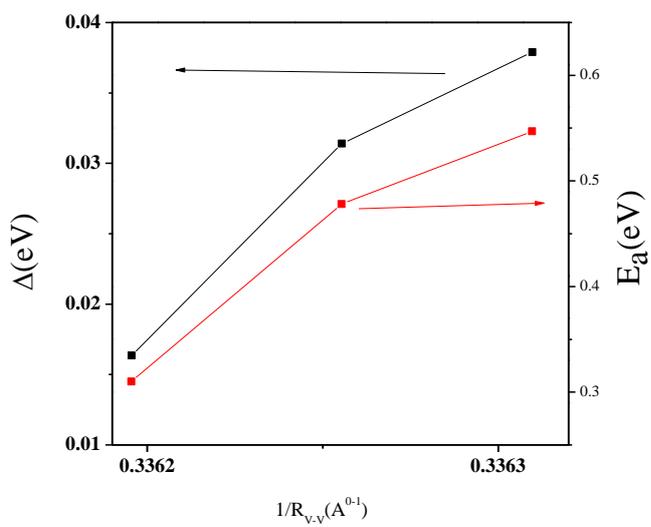

Figure 14